\documentclass[10pt, conference]{IEEEtran}

\ifCLASSOPTIONcompsoc
  \usepackage[nocompress]{cite}
\else
  \usepackage{cite}
\fi
\def\BibTeX{{\rm B\kern-.05em{\sc i\kern-.025em b}\kern-.08em
    T\kern-.1667em\lower.7ex\hbox{E}\kern-.125emX}}

\ifCLASSINFOpdf
  \usepackage[pdftex]{graphicx}
  \graphicspath{{figures/}}
  \DeclareGraphicsExtensions{.pdf,.jpeg,.png}
\else
\usepackage[dvips]{graphicx}
  \graphicspath{{figures/}}
  \DeclareGraphicsExtensions{.eps}
\fi

\usepackage{enumitem}
\usepackage{multirow}
\usepackage{xcolor}
\usepackage[skip=5pt]{caption}

\ifCLASSOPTIONcompsoc
  \usepackage[caption=false,font=footnotesize]{subfig}
\else
  \usepackage[caption=false,font=footnotesize]{subfig}
\fi

\usepackage{verbatim}
\usepackage{soul}
\usepackage{xcolor}

\usepackage{url}

\usepackage{listings}
\usepackage{color}

\definecolor{mygreen}{rgb}{0,0.6,0}
\definecolor{mygray}{rgb}{0.5,0.5,0.5}
\definecolor{mymauve}{rgb}{0.58,0,0.82}

\usepackage{fancyhdr}

\begin{document}
\bstctlcite{IEEEexample:BSTcontrol}
\renewcommand{\arraystretch}{1.2}

\lstdefinestyle{interfaces}{
  float=tp,
  floatplacement=tbp,
  abovecaptionskip=-5pt
}

\title{Union: An Automatic Workload Manager for Accelerating Network Simulation
}

\author{\fontsize{11}{11}\selectfont
{
Xin Wang,\IEEEauthorrefmark{1}
Misbah Mubarak,\IEEEauthorrefmark{2}\IEEEauthorrefmark{3} 
Yao Kang,\IEEEauthorrefmark{1}
Robert B. Ross,\IEEEauthorrefmark{2} 
Zhiling Lan\IEEEauthorrefmark{1}
}
\vspace{1.6mm}\\
\fontsize{10}{10}\selectfont\itshape
\IEEEauthorrefmark{1}Department of Computer Science, 
Illinois Institute of Technology, Chicago, IL 60616, USA\\
\fontsize{9}{9}\selectfont\ttfamily\upshape
\{xwang149, ykang17\}@hawk.iit.edu, lan@iit.edu

\vspace{1.2mm}\\
\fontsize{10}{10}\selectfont\rmfamily\itshape
\IEEEauthorrefmark{2} Mathematics and Computer Science Division,
Argonne National Laboratory, Argonne, IL 60439, USA\\
\fontsize{9}{9}\selectfont\ttfamily\upshape
\{mmubarak, rross\}@mcs.anl.gov
}


\maketitle

\thispagestyle{fancy}
\lhead{}
\rhead{}
\chead{}
\lfoot{\footnotesize{
2020 IEEE International Parallel and Distributed Processing Symposium (IPDPS)}}
\rfoot{}
\cfoot{}
\renewcommand{\headrulewidth}{0pt}
\renewcommand{\footrulewidth}{0pt}

\begin{abstract}
With the rapid growth of the machine learning applications, the workloads of future HPC systems are anticipated to be a mix of scientific simulation, big data analytics, and machine learning applications. 
Simulation is a great research vehicle to understand the performance implications of co-running scientific applications with big data and machine learning workloads on large-scale systems. 
In this paper, we present Union, a workload manager that provides an automatic framework to facilitate hybrid workload simulation in CODES. 
Furthermore, we use Union, along with CODES, to investigate various hybrid workloads composed of traditional simulation applications and emerging learning applications on two dragonfly systems. 
The experiment results show that both message latency and communication time are important performance metrics to evaluate network interference. 
Network interference on HPC applications is more reflected by the message latency variation, whereas ML application performance depends more on the communication time. 
\end{abstract}


\footnotetext[1]{Zhiling Lan's current affiliation is University of Illinois Chicago, and her current contact is \ttfamily\upshape{zlan@uic.edu}.}
\footnotetext[3]{Misbah Mubarak's current affiliation is Amazon Web Services, and her current contact is \ttfamily\upshape{mimubara@amazon.com}.}

\section{Introduction}
Recently, high-performance computing (HPC) portfolio has diversified beyond the traditional simulation focus to include significant amount of activities employing machine learning and data analytics techniques.
The community is embracing machine learning (ML) and other artificial intelligence (AI) techniques for countless pursuits, from driving groundbreaking scientific discoveries to protecting national security. 
Extreme-scale supercomputers have been proven suited to these emerging applications\cite{ben2018demystifying}\cite{you2018imagenet}.  
The convergence of HPC, AI, and data analytics is underway to better leverage the investment of supercomputers.
For example, current and up-coming supercomputers such as Summit at Oak Ridge, Frontera at TACC, and Aurora at Argonne are all built for supporting traditional scientific simulations and emerging AI applications.
Moreover, applying AI and data analytics on HPC systems can dramatically increase the value and utilization of resources, hence boosting the productivity of the HPC systems.
Together, HPC and AI will accelerate scientific discoveries that we haven't yet dream of.


While there may be drastic behavioral difference between scientific applications and data analysis applications using ML, both of them have significant communication requirements. 
For example, the gradient aggregation communication for a deep learning application achieves 1.7 GB/s per node \cite{mathuriya2018cosmoflow}, and a typical HPC application (MILC) issues hundreds of thousands of nonblocking communication of 10 KB--364 KB messages.
These requirements place a heavy burden on the interconnect network of supercomputers. 

The ever-increasing need for higher bandwidth and higher message rate has driven the design of low-diameter interconnect topologies like variants of dragonfly (1D \cite{kim2008technology}, 2D \cite{faanes2012cray}, D+ \cite{flajslik2018megafly}).  
As these hierarchical networks become increasingly dominant, application performance variability becomes a serious issue \cite{chunduri2017run, jokanovic2015quiet, kambadur2012measuring}. 
Unfortunately, \emph{Little work has been conducted to understand performance implications of co-running scientific applications with big data analysis on dragonfly systems.} 
We need to study how different interconnect technologies affect workload performance, and we need to understand how conventional scientific applications interact with emerging big data and machine learning applications at the underlying interconnect level. 
Moreover, given the potential diversity of interconnect networks in the future, there is an even greater need for tools enabling extensive what-if analysis when exploring the design spaces of various application-system configurations. 

While real-world experiment is the best way to evaluate hybrid workloads on a target system, it is unrealistic to fully rely on experiments for permanence analysis, especially when researching on various system designs. \emph{Modeling and simulation} provides a powerful alternative to experiments for designing and evaluating system behaviors. 
Moreover, it is an indispensable tool for exploring various design alternatives (e.g., diverse workloads on different system configurations). 

There are several well-known system modeling toolkits in HPC \cite{carothers2002ross, mubarak2017enabling, rodrigues2011structural, jiang2013detailed}. CODES is an open-source, community-built toolkit which provides a set of flit-level HPC interconnect models for users to simulate different network designs, and ROSS serves as its underlying event-driven simulation framework \cite{mubarak2017enabling}. 
In this study, we will use CODES to analyze heterogeneous workloads on various dragonfly systems. 

Currently, CODES supports both trace-based simulation and skeleton simulation using SWM \cite{thompson2014scalable}.
In \emph{trace-based simulation}, the application traces 
are collected by executing the application on real system
, thus the accuracy of simulation is guaranteed. 
However, trace-driven simulation has the drawbacks of limited scalability and huge memory footprint caused by large trace size and intensive communication traffic. 
Moreover, trace-based analysis cannot be easily scaled to a different number of processors. 

\emph{Skeleton simulation} becomes popular for large-scale simulation \cite{jain2016evaluating}\cite{mubarak2019evaluating}. Skeleton is a curtailed version of the full application such that expensive computation is replaced with delay models, which significantly reduce simulation cost without sacrificing simulation accuracy. 
However, skeleton simulation using SWM is complex and time consuming, requiring significant efforts to develop skeletons and to integrate them in CODES.

By far, there is no feasible toolkit available in CODES for us to perform large-scale simulations with intensive hybrid workloads.
In this work, we develop \emph{Union}, a workload manager to facilitate hybrid workload simulation in CODES.
Users only need to write simple English instructions to describe an application. Union automatically translates these instructions into a skeleton and coordinates the skeleton generation in CODES.
Moreover, we use Union, along with CODES, to investigate hybrid workloads with both ML applications and traditional scientific applications on two 8,488-node HPC systems.
The experiments results reveal several key findings:

\begin{itemize}
\item Message latency is a reliable metric to reflect network interference. 
Application with intensive communication patterns suffers less slowdown in message latency than communication non-intensive ones. 
Placing communication-intensive application into separate groups helps confine their messages within the assigned groups, hence mitigating its interference to other applications. 
\item 
The increase in the message latency affects HPC applications more than ML applications in term of communication time, implying that the ML application has better ability to \textit{absorb} the message delays \cite{de2019mitigating}. 
\item 
In our system setup, applications achieve better performance the on 2D dragonfly system than on the 1D dragonfly system because 2D dragonfly system offers more global and local links to mitigate network congestion.
\end{itemize}

The remainder of this paper is organized as follows:
Section II introduces coNCePTuaL and CODES, 
Section III-IV describes Union and our methodology,
Section V validates Union,
Section VI presents the experimental results and analysis, 
Section VII discusses related topics,
Section VIII presents the related works, 
and Section IX draws some conclusions from the information presented in this paper.

\section{Background}



\begin{figure}[ht]
\caption{A sample coNCePTuaL code for Ping-Pong test.}
  \begin{lstlisting}[frame=single, escapechar=\&]
1  # A ping-pong latency test written in coNCePTuaL
2  Require language version "1.5".
3  
4  # Parse command line.
5  reps &\textbf{is}& "Number of repetitions" &\textbf{and}& &\textbf{comes}& &\textbf{from}& "--reps" &\textbf{or}& "-r" &\textbf{with}& &\textbf{default}& 1000.
6  msgsize &\textbf{is}& "Message size of bytes to transmit" &\textbf{and}& &\textbf{comes}& &\textbf{from}& "--msgsize" &\textbf{or}& "-m" &\textbf{with}& &\textbf{default}& 1024.
7  
8  &\textbf{Assert}& that "the latency test requires at least two tasks" &\textbf{with}& num_tasks>=2.
9  
10 # Perform the test.
11 &\textbf{For}& reps &\textbf{repetitions}& {
12    &\textbf{task}& 0 &\textbf{resets its counters then}&
13    &\textbf{task}& 0 &\textbf{sends}& &\textbf{a}& msgsize &\textbf{byte}& &\textbf{message}& to &\textbf{task}& 1 &\textbf{then}&
14    &\textbf{task}& 1 &\textbf{sends}& &\textbf{a}& msgsize &\textbf{byte}& &\textbf{message}& to &\textbf{task}& 0 &\textbf{then}&
15    &\textbf{task}& 0 &\textbf{logs}& &\textbf{the}& msgsize &\textbf{as}& "Bytes" &\textbf{and}& &\textbf{the}& &\textbf{median}& &\textbf{of}& elapsed_usecs/2 &\textbf{as}& "1/2 RTT (usecs)"
16 } &\textbf{then}&
17    &\textbf{task}& 0 &\textbf{computes}& &\textbf{aggregates}&
\end{lstlisting}
\label{fig:conc-code} 
\end{figure}

\subsection{coNCePTuaL}
coNCePTuaL (Network Correctness and Performance Testing Language) \cite{pakin2007design} is a domain-specific specification language dedicated to help measure the performance and correctness of networks. 
coNCePTuaL is featured with primitives that are frequently used in parallel applications, which can be used to not only describe communication behavior but also simulate computation and I/O.

coNCePTuaL contains two main components: a \emph{domain-specific language (DSL)} and a \emph{compiler}. 
The domain-specific language is expressly developed for writing network benchmarks.
coNCePTuaL provides a keyword-heavy syntax that reads like an English-language description.
Figure \ref{fig:conc-code} shows an example ping-pong benchmark written in coNCePTuaL language, with keywords shown in bold. 
A complete benchmark includes command-line parsing, execution, timing, and statistics logging.
It emphasizes the communication pattern, encapsulates other routine activities such as initialization of messaging libraries, allocation of data structures, variable declaration, and statistics recording.

The coNCePTuaL compiler contains: a lexer converting coNCePTuaL source code into a token list; a parser converting the token list into an abstract syntax tree (AST); and a code generator converting the AST into low-level code including calls to a messaging library~\cite{pakin2004conceptual}. 
The compiler supports a variety of code generators, the most commonly used is the C + MPI generator that produces a C code with calls to an MPI library for message passing.

A salient feature of coNCePTuaL is its built-in functions to support various virtual topologies in application communication such as n-ary trees, meshes, tori, and k-nomial trees. 
These functions can significantly reduce the manual effort to implement complex communication behaviors.
Thus, we use coNCePTuaL instead of common workflow language like C to write applications for various network performance studies.



\begin{figure}[ht]
\centering
\includegraphics[width=0.43\textwidth]{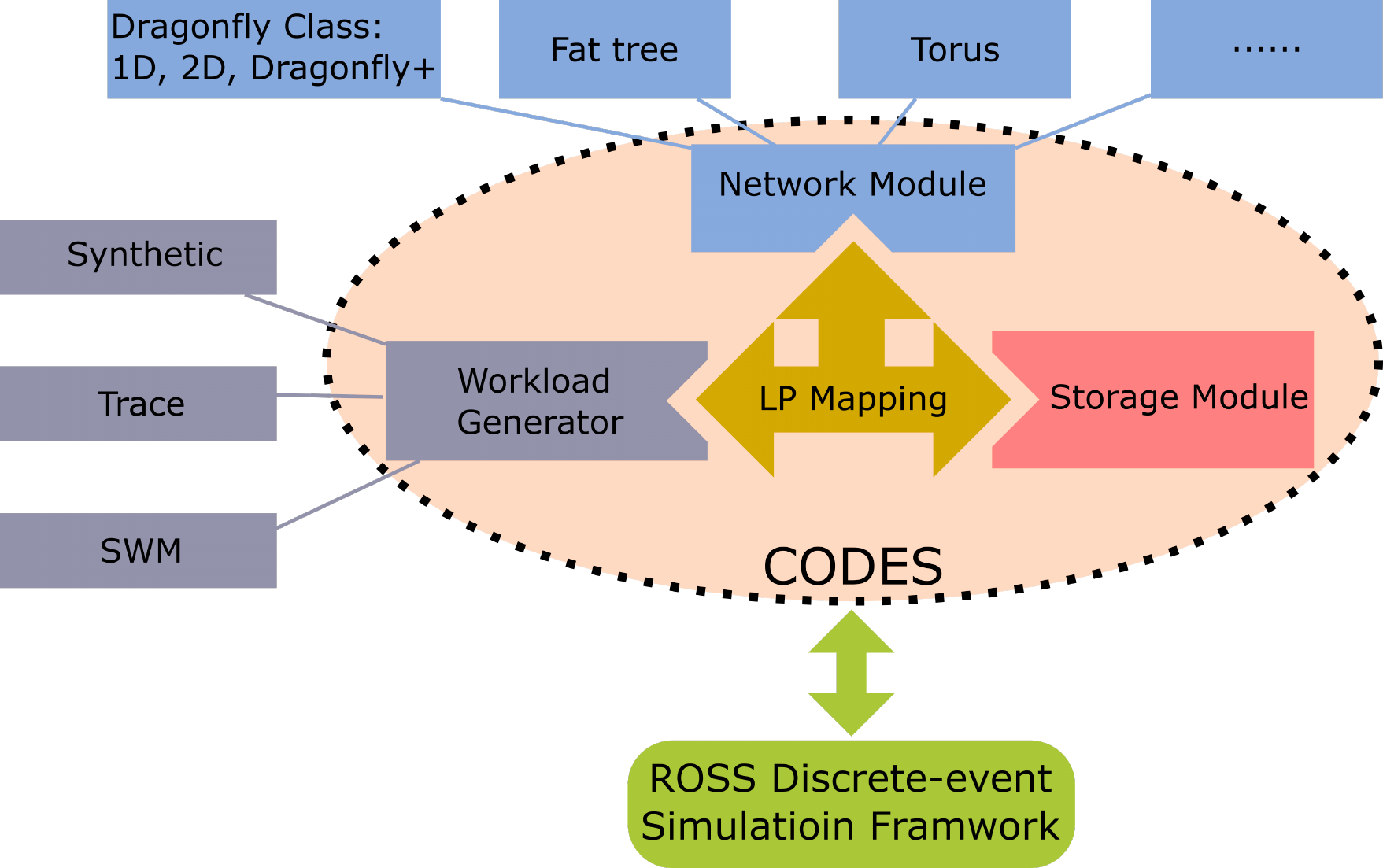}
\caption{Overview of CODES}
\label{fig:codes}
\end{figure}

\subsection{CODES}

CODES (Enabling CO-Design of Exascale Storage Systems) is a parallel event-driven simulator, which enables packet-level, high-fidelity simulation to explore the design of large-scale storage and network architectures \cite{mubarak2017enabling}.
Figure \ref{fig:codes} illustrates the main components in the CODES framework including a workload generator, a network module, and a storage module.
CODES is built upon the Rensselaer Optimistic Simulation System (ROSS), a discrete event simulation framework that allows simulations to be run in parallel. 
Network simulation is one of the key features in CODES. The network module provides an abstraction layer for various network topologiy models to plug in, including dragonfly class, Torus, Fat-Tree, Slim Fly, and many more \cite{wang2018trade, mubarak2017quantifying, Yang:ICPADS:2016}.


\begin{figure*}[ht]
\centering
\includegraphics[width=0.90\textwidth]{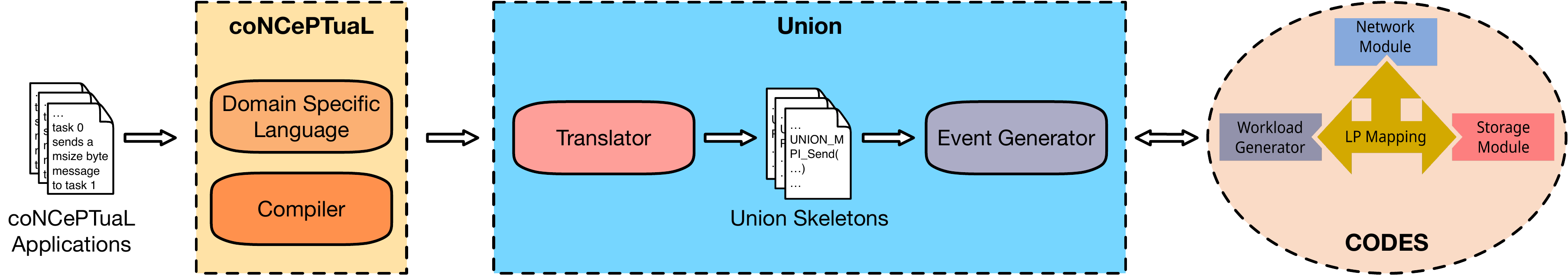}
\caption{Diagram of in situ simulation framework with Union}
\label{fig:cowg_arch}
\end{figure*}

The CODES workload generator supports abstractions that allow I/O and network workloads from various sources to drive the underline network and storage models. 
The sources of network workloads currently support including synthetic workloads, application traces, and SWM skeletons \cite{mubarak2019evaluating}. 
Synthetic network workloads include uniform random traffic and nearest-neighbor traffic. 
Application traces are MPI traces generated by the SST DUMPI library \cite{rodrigues2011structural}.
SWM skeleton is executed as in situ workloads with the CODES simulation. 
Argobots \cite{seo2017argobots}, a lightweight threading and tasking tool, is used for coordinating the execution of the SWM skeleton and CODES.
During simulation, CODES creates a separate light-weight thread to represent each process in the skeleton program. 
Each skeleton thread executes the skeleton code and issues communication calls. 
Instead of initiating message exchange, skeleton threads yield to CODES for processing the communication events. 
After proceeding all the events, CODES then yields to the skeleton threads for more events. 
Argobots handles the synchronization of threads.

The workflow to develop a new SWM skeleton is described as follows: 
(i) manually transform a full application to a skeleton by replacing expensive computation with delay models for CODES to estimate computation time and eliminating unnecessary variable assignments to reduce memory footprint,
(ii) manually modify the in situ workload generator in CODES to register the new SWM skeleton,
and (iii) recompile CODES.

In summary, current SWM workflow is cumbersome with tedious and error-prone human effort. 
In this work, we develop Union, an in situ workload manager for CODES, which handles the aforementioned workflow automatically.

\section{Union Design}

\subsection{Union Features}

\begin{table}[t]
\centering
\captionsetup{justification=centering}
\caption{Comparison between different workload generating frameworks in CODES}
\label{tab:comparison}
\begin{tabular}{l|c|c|c}
\cline{1-4}
\hline\hline
Features   & Trace Replay & SWM & Union  \\ \cline{1-4}
Trace collection 				
				 &  Yes  &  No  &  No  									\\
Memory foodprint   	 
				 &  Large  &  Small  &  Small  						\\
Scaling application size	 
				 &  Re-tracing  &  Yes  &  Yes  					\\
Automatic Skeletonization 
				 &  N/A  &  No  &  Yes  								\\
Integration to CODES
				 &  Easy  &  Human  &  Automated 				 \\
Validation w/ new hardware
				 &  Re-tracing  &  Re-written  &  Easy 			 \\				
\hline\hline
\end{tabular}
\end{table}

Table \ref{tab:comparison} summarizes the comparison between different workload generating mechanisms in CODES. 
Union is considered as a better solution with the following features:
\begin{itemize}
\item \emph{Unification}: the applications have unified syntax and execution flow to support automatic post-processing. A domain-specific language is well-suited for this purpose.
\item \emph{Automation}: the skeleton is automatically generated from application, which reduces human effort and avoids human errors. 
\item \emph{Effortlessness}: integrating new applications to simulation framework takes almost no human effort. Application programmer does not need to have prior knowledge about the implementation of the simulator.
\item \emph{Deployability}: validation of simulation results with new hardware is straightforward by running the full application on the new machine, since the skeleton is directly derived from the full application.
\end{itemize}


\subsection{System Architecture}
Union contains two main components, a \emph{translator} that automatically translates coNCePTuaL applications into skeletons, and an \emph{event generator} that emits communication events from skeletons to CODES as an in situ workloads. 

Figure \ref{fig:cowg_arch} illustrates the high-level architecture of the in situ simulation framework with Union.
The translator takes applications written in coNCePTuaL language as inputs, collaborates with coNCePTuaL compiler to build Union skeletons. Figure \ref{fig:cowg_object} presents the code snippet of a Union skeleton generated from the Ping-Pong program shown in Figure \ref{fig:conc-code}.
The event generator is an abstraction layer that allows Union skeletons to be used as pluggable in situ workloads for CODES simulation framework. 
The event generator unifies the structure of Union skeletons, and provides message passing API to work in conjunction with the  workload generator in CODES for extracting communication events from Union skeletons.

\subsection{Implemetation}
\begin{figure}[h]
\centering
\includegraphics[width=0.47\textwidth]{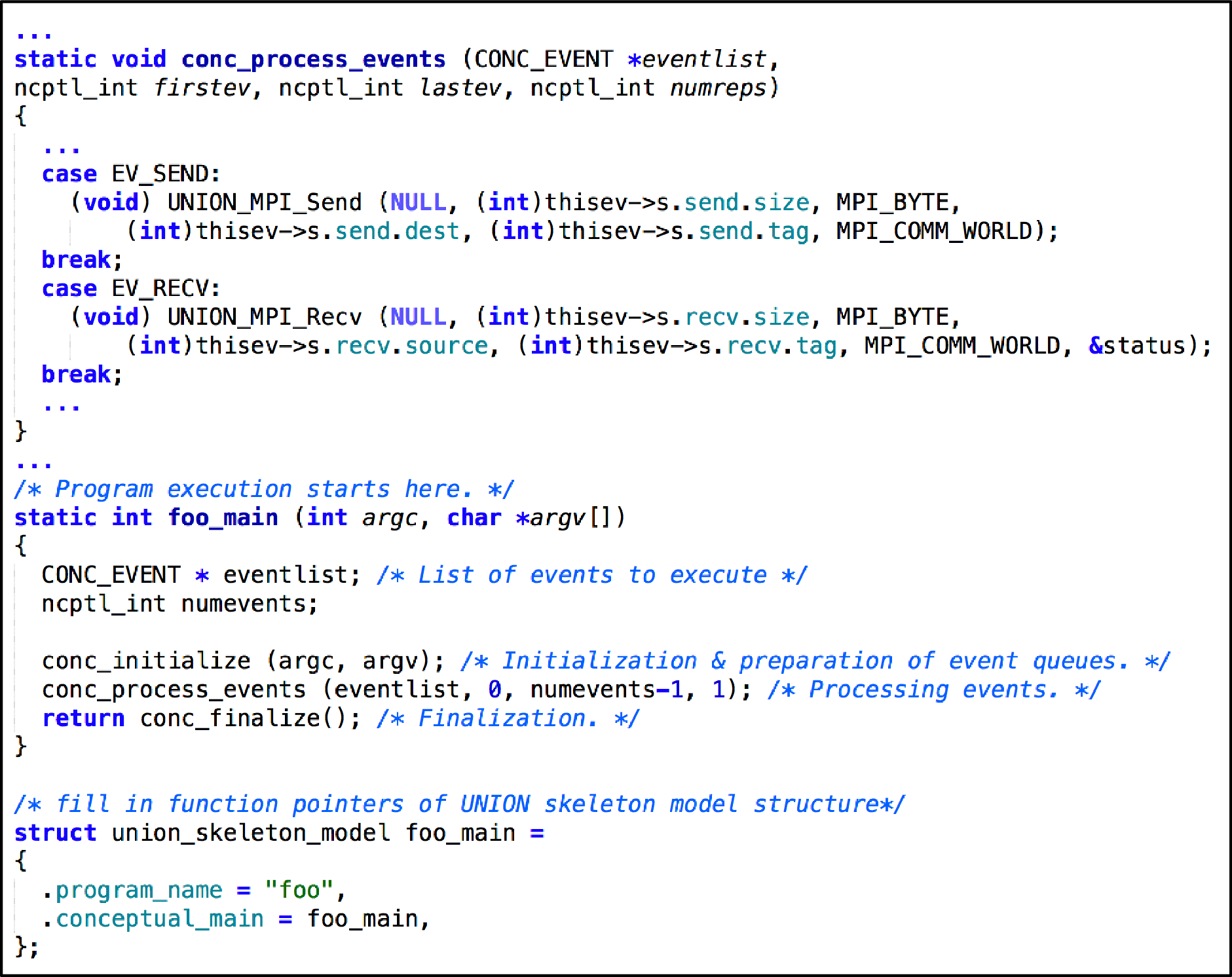}
\caption{Structure that defines a Union skeleton object.}
\label{fig:cowg_struct}
\end{figure}

Union maintains a list of available skeleton objects defined in a data structure as shown in Figure \ref{fig:cowg_struct}. 
A skeleton object simply contains the name of the program and a declaration of the main function. 

\begin{figure}[h]
\centering
\includegraphics[width=0.47\textwidth]{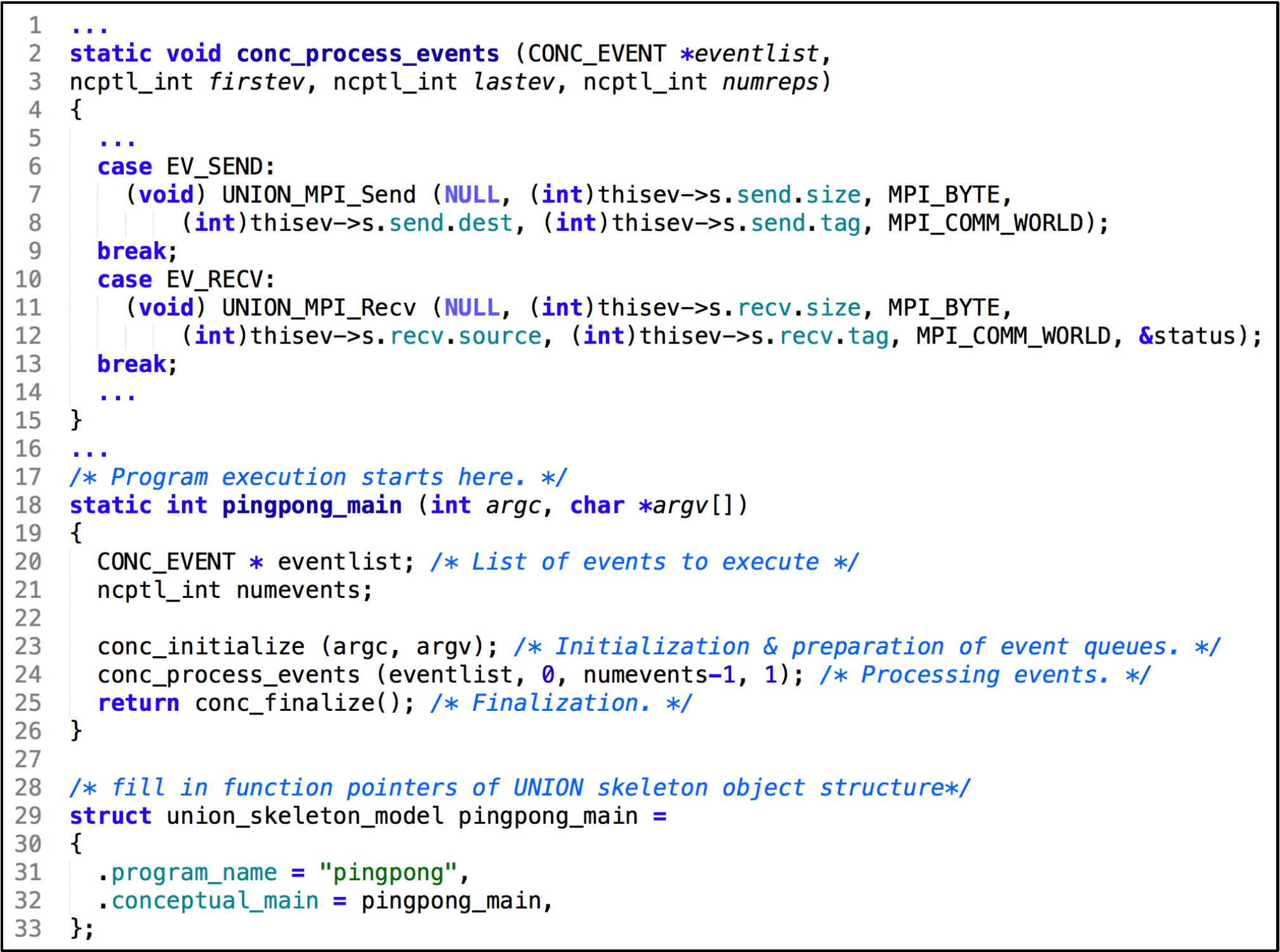}
\caption{An example code snippet of a Union skeleton generated from the Ping-Pong test in Figure \ref{fig:conc-code}. 
Line 23 handles parsing of command line (line 5-8 of Figure \ref{fig:conc-code}) and initialization of event queues (line 11-17 of Figure \ref{fig:conc-code}).
Line 24 processes all the events of the ping-pong application shown in Figure \ref{fig:conc-code}. Line 6-13 intercept and translate the communication operations (line 13 \& 14 of Figure \ref{fig:conc-code}) to Union message passing interfaces.
Some portions of the code are skipped due to space limit.}
\label{fig:cowg_object}
\end{figure}

The translator inherits the functions from the general C backend compiler in coNCePTuaL, which is used to transform the abstract syntax tree generated from a coNCePTuaL source code to a Union skeleton. 
The translator takes three steps to add a new application into the framework. 
The first step is the initialization of a Union skeleton object.
The translator constructs a benchmark object by filling the name and main function pointers as shown in Figure \ref{fig:cowg_object} (line 28-33), and adding the object to the available skeleton object list.
The second step is skeletonization. The translator changes all communication buffers to null to reduce memory footprint. 
In terms of computation, coNCePTuaL encapsulates the computation into a tight spin-loop and sleeps for a given length of time, thus we add a UNION\_Compute() method that instructs the simulator to account for the computation delay. 
The third step is to intercept communication operations. The translator forces all communication function calls to use the Union message passing API. For example, \texttt{MPI\_Send()} calls are converted to \texttt{UNION\_MPI\_Send()} as shown in Figure \ref{fig:cowg_object} (line 6-13).

The event generator in Union declares message passing interface in the format of \texttt{UNION\_MPI\_X}. We add a pluggable workload module into CODES workload generator to hold the actual implementation of these operations, such that the messages from Union skeletons can be emitted as simulation events in CODES.

Union works in harmony with the concurrent workload support and the flexible rank-to-node mapping. We can co-run multiple large-scale skeletons with any predefined rank-to-node allocation and routing police in one simulation.

\section{Evaluation Methodology}
In this section, we describe the methodology for the hybrid workload analysis. 
We conduct large-scale simulation study of hybrid HPC and ML workloads on 8,448-node systems. 

\begin{table*}[t]
\centering
\captionsetup{justification=centering}
\caption{Configuration of two HPC systems.}
\label{tab:sys}
\begin{tabular}{c|c|c|c|c|c|c|c}
\hline\hline
Topology     & Radix & \#Groups & \#Routers/Group & \#Nodes/Router & \#Nodes/Group  & \#Global Channel/Router & System Size\\ \hline 
1D dragonfly & 48    & 33        & 32               & 8               & 256        & 4             & 8448                        \\ 
2D dragonfly & 48    & 22        & 96               & 4               & 384        & 7             & 8448                        \\ \hline\hline
\end{tabular}
\end{table*}

\subsection{System Configuration}

We study two different HPC networks: 1D and 2D dragonfly. 
Dragonfly network has a 2-level hierarchical design. For both 1D and 2D networks, the system's compute nodes are divided into several identical groups, which are all-to-all connected. 
Within a group, 1D and 2D dragonfly have different configurations. 
In 1D dragonfly, the routers within the same group are all-to-all connected. This topology is expected to be used in the upcoming exascale systems.
In 2D dragonfly, each group has 96 routers arranged in a $6 \times 16$ matrix, the routers share the same row or column are all-to-all connected. 
This topology is adopted by Cori at NERSC and Theta at Argonne.
We use 48-ports routers for both networks. The configurations of the 1D dragonfly and 2D dragonfly are given in Table \ref{tab:sys}. 
The terminal, local and global link bandwidth are set to be 16 GiB/s, 4.69 GiB/s and 5.25 GiB/s respectively. 

\subsection{Hybrid Workloads}
\label{sec:wkld}
In this study, we investigate three hybrid workloads composed of multiple HPC applications and ML applications. HPC applications include a synthetic nearest neighbor application and three SWM skeletons. We build two ML skeleton applications by using Union. The details are listed below:

\textbf{Cosmoflow}. Distributed machine learning algorithms are featured with periodic Allreduce calls to gather gradients from multiple worker nodes and broadcast summation result to them. This application captures this feature by iteratively issuing Allreduce calls with a predefined compute time interval. 
It is configured as a 1,024-rank job that issues 28.15 MiB Allreduce messages every 129 ms as described in \cite{mathuriya2018cosmoflow}.

\textbf{AlexNet}. AlexNet is the name of a convolutional neural network, designed by Alex Krizhevsky. 
We collect the communication traces from a 512-node execution of AlexNet. Since Horovod \cite{sergeev2018horovod} is used as the distributed training framework, we observe lots of small 4-byte and 25-byte negotiation messages before each gradient update. Each gradient update contains several Allreduce calls transmitting a total of 235 MiB messages.
We model the traced communication patterns as well as the computation interval, and create this application to represent the communication behavior in AlexNet. 

\textbf{Nearest Neighbor (NN)}. This synthetic pattern represents a common kernel in multiple scientific applications including algebraic multiGrid solver (AMG), Hardware Accelerated Cosmology Code (HACC), etc. 
The processes are formed into a 3D Cartesian grid. In each iteration, every process communicates with neighbors in each dimension. In this study, it is configured with 512 ranks, transmitting 128 KiB messages with nonblocking send and receive.

\textbf{MILC}. 
MILC is developed by the MIMD Lattice Computation (MILC) collaboration to study quantum chromodynamics (QCD).
It performs simulations of four dimensional SU(3) lattice gauge theory. The SWM of MILC extracts the communication pattern of MILC. It is configured with 4,096 ranks, each rank issues nonblocking send and receive messages of size 486 KiB to communicate with neighbors.  

\textbf{Nekbone}. 
Nekbone is a mini-app derived from the computational fluid dynamics code Nek5000.
Nekbone solves a standard Poisson equation using a conjugate gradient iteration with a simple preconditioner.
The SWM of Nekbone is configured with 2,197 ranks and performs a large number of MPI collective operations with small 8-byte messages. It uses nonblocking send and receive to transmit messages with various sizes from 8 bytes to 165 KiB.

\textbf{LAMMPS}.  
LAMMPS is a classical molecular dynamics simulation code designed to run efficiently on parallel computers.
The SWM version of LAMMPS is configured with 2,048 ranks. It uses Allreduce calls with small messages, and blocking send and nonblocking receive with various message sizes from 4 bytes to 135 KiB.

Table \ref{tab:workloads} lists the hybrid workloads analyzed in this study. 
In Workload1, uniform random (UR) synthetic background traffic is configured with 4,096 ranks, each rank sending 10 KiB message at 1 ms interval.

We first collect the performance data of baseline cases that each application is independently simulated with no other jobs sharing the network. Then we simulate the three mixed workloads, collect performance metrics of each application, and compare them with the baseline cases. 
Each aforementioned simulation is conducted with 6 different combinations of job placement policies and routing mechanisms.

\begin{table}[ht]
\centering
\captionsetup{justification=centering}
\caption{Hybrid HPC and ML workloads}
\label{tab:workloads}
\begin{tabular}{c|c|c|c}
\hline\hline
Workload  & ML Skeletons     & SWM Skeletons          & Synthetic \\ 
\hline
Workload1 & Cosmoflow, AlexNet & LAMMPS, NN          & UR         \\ 
\hline
Workload2 & Cosmoflow, AlexNet & LAMMPS, MILC, NN  &            \\
\hline
Workload3 & Cosmoflow, AlexNet & Nekbone, MILC, NN &            \\ 
\hline\hline
\end{tabular}
\end{table}


\begin{figure*}[ht]
\centering
\includegraphics[width=0.90\textwidth]{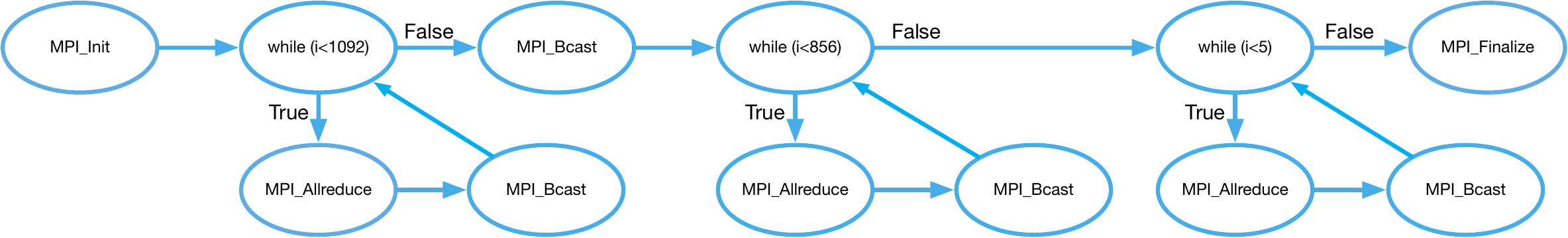}
\caption{Control flow graph of AlexNet. Both the application and the skeleton follow the exact control flow.}
\label{fig:control_flow}
\end{figure*}

\subsection{Job Placement and Routing}
In this study, we investigate three job placement policies.
\begin{itemize}
\item \textbf{Random Nodes (RN)} selects compute nodes for each job completely randomly from the entire system. Compute nodes that connect to the same router tend to be assigned to different jobs. 
\item \textbf{Random Routers (RR)} assigns each job a random selection of routers, compute nodes connected to that router are assigned consecutively. This scheme helps prevent contention within a router among different jobs. 
\item \textbf{Random Group (RG)} assigns each job a random selection of groups, nodes in the groups are assigned consecutively. This method tends to place different application processes into different groups.
\end{itemize}

We study two commonly used routing algorithms.
\begin{itemize}
\item \textbf{Minimal Routing (MIN)} routes a packet along the minimal path from source to destination. 
Minimal routing can guarantee the minimum hops a packet traverses.
\item \textbf{Adaptive Routing (ADP)} selects the path taken by a packet based on the congestion situation on minimal and non-minimal paths. When a non-minimal path is chosen, the packet will be minimally routed to a random intermediate router, then minimally forwarded to its destination. Adaptive routing is designed to avoid hotspots and to balance network traffic.
\end{itemize}

\subsection{Performance Metrics}
The performance metrics we analyze including \emph{communication time}, \emph{message latency}, and \emph{message amount on routers}. 
Communication time is defined as the portion of process runtime used for sending and receiving messages. 
Message latency is the time that each message spends to reach its destination from the source. 
The communication time and message latency are used to quantify network interference. Each process records minimal, average and maximum message latency among all the messages they receive. 
We implement a packet counter for each application in the router module of CODES. On each router, this counter records the total packets it receives for each application during a configurable time window. Knowing the packet size and the time window size, we can easily calculate the data arrival rate on this router. All following experiments in this study use a 0.5 ms time window.

The experiments are conducted on the Bebop machine at Argonne National Laboratory \cite{Bebop}. 
Bebop is equipped with 1,024 nodes, including 664 Intel Broadwell nodes and 352 Knights Landing nodes. 
Each Broadwell nodes contains a 36-core processor with 128 GB of DDR4 RAM. 
All of our experiments use the optimistic parallel model in CODES/ROSS, and are executed on 4 Broadwell nodes.
The average simulation runtime is approximately 5 hours. 

The simulation replays the hybrid workloads composed of ML and HPC skeletons, both are communication intensive. 
The observed peak message injection rate during simulation is 160 TiB/s. 
Without the in situ workload generation framework using Union, such large-scale simulations can not completed within a reasonable time.

\begin{table}[h]
\centering
\captionsetup{justification=centering}
\caption{AlexNet - MPI event count}
\label{tab:correct1}
\begin{tabular}{l|l|l}
\hline\hline
Function & Application & Union Skeleton \\ \hline
\texttt{MPI\_Init}           	& 512			& 512      	\\ 
\texttt{MPI\_Bcast}       	& 1969		& 1969		\\ 
\texttt{MPI\_Allreduce}  	& 1958  		& 1958		\\ 
\texttt{MPI\_Finalize}		& 512  		& 512			\\ 
\hline\hline
\end{tabular}
\end{table}

\begin{table}[h]
\centering
\captionsetup{justification=centering}
\caption{AlexNet - Bytes transmitted by each rank}
\label{tab:correct2}
\begin{tabular}{l|l|l}
\hline\hline
Rank & Application & Union Skeleton 	\\ \hline
0           		& 6.33e11					& 6.33e11     					\\ 
1 to 511		& 2.47e8 + 6.33e11	& 2.47e8 + 6.33e11		\\ 
\hline\hline
\end{tabular}
\end{table}

\section{Union Validation}

Skeleton correctness is of utmost importance when applying skeleton-driven approach. In order to use a skeleton in place of an application, the runtime behavior of the skeleton has to match the application's behavior both in terms of control flow and communication pattern.
Here, control flow indicates the order in which instructions and function calls of a program are executed. 
With respect to communication pattern, we mainly focus on matching the data transmitted per MPI rank.

Here, we present the validation result of AlexNet listed in Section \ref{sec:wkld}.
Figure \ref{fig:control_flow} presents the control flow of AlexNet extracted from both the application and the skeleton. 
We calculate the number of times an MPI event occurs during the executions of the application and the skeleton, as shown in Table \ref{tab:correct1}. 
The MPI events are grouped by the function name and the count is shown for each function.
For each MPI function call in Table \ref{tab:correct1}, we observe that the number of events extracted from the application and the skeleton are equal. 
This demonstrates that the skeleton has correct control flow. 

\begin{figure*}[ht]
\centering
\subfloat[LAMMPS]{
   \includegraphics[width=0.32\textwidth]{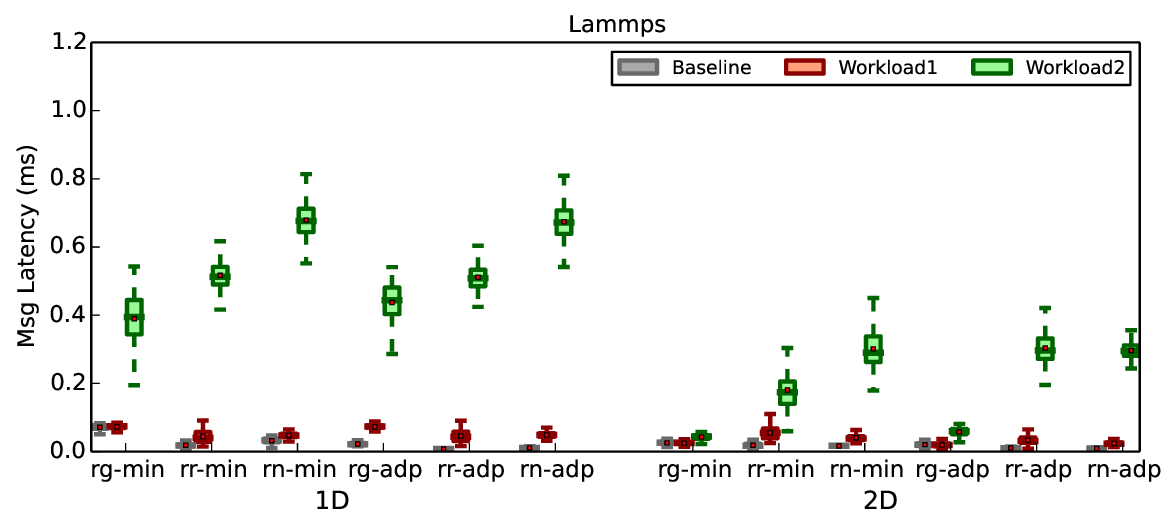}
}
\subfloat[Nekbone]{
   \includegraphics[width=0.32\textwidth]{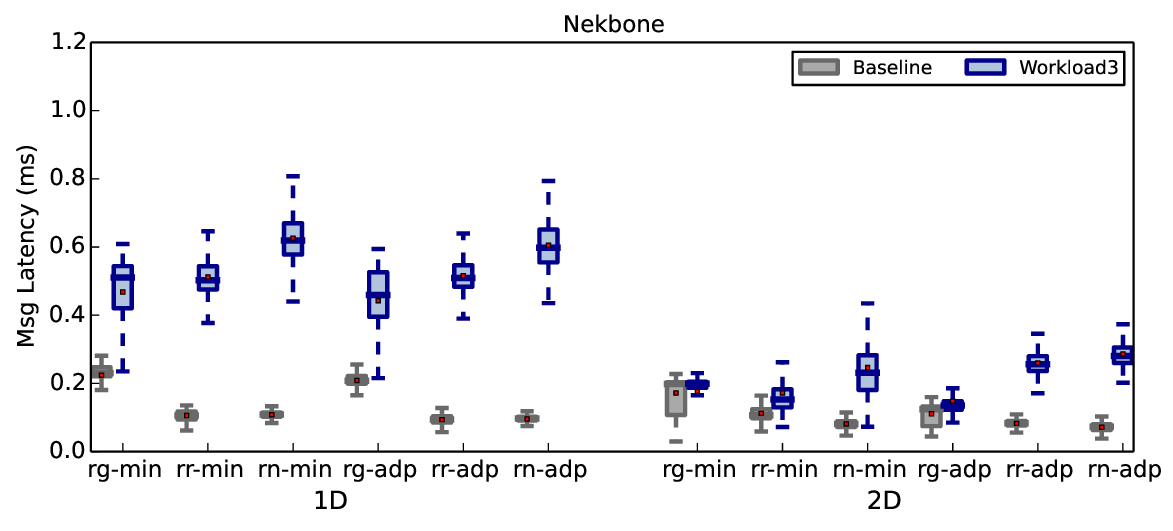}
}
\subfloat[MIILC]{
   \includegraphics[width=0.32\textwidth]{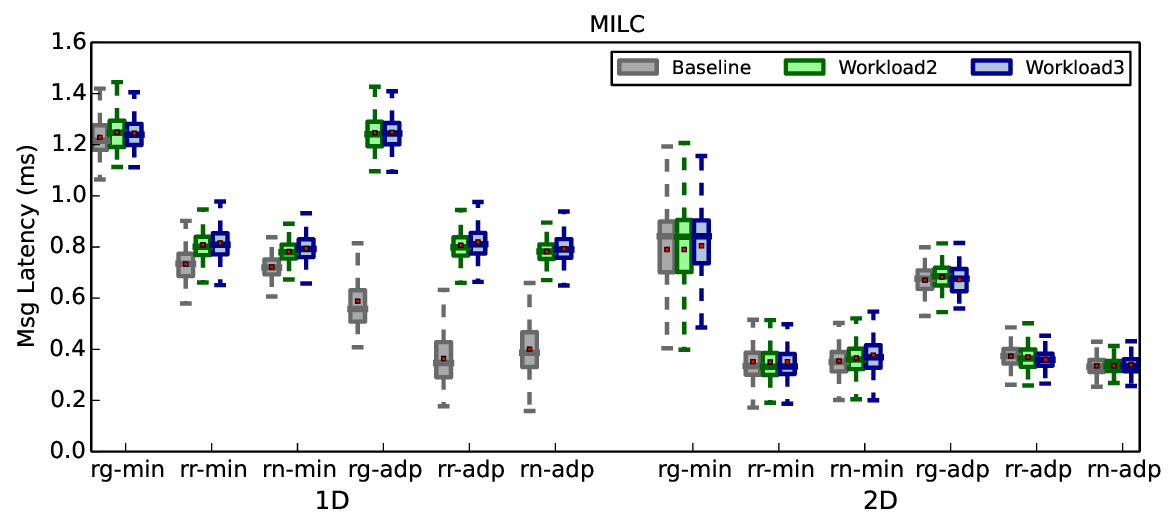}
}\\
\vspace{-8pt}
\subfloat[AlexNet]{
   \includegraphics[width=0.48\textwidth]{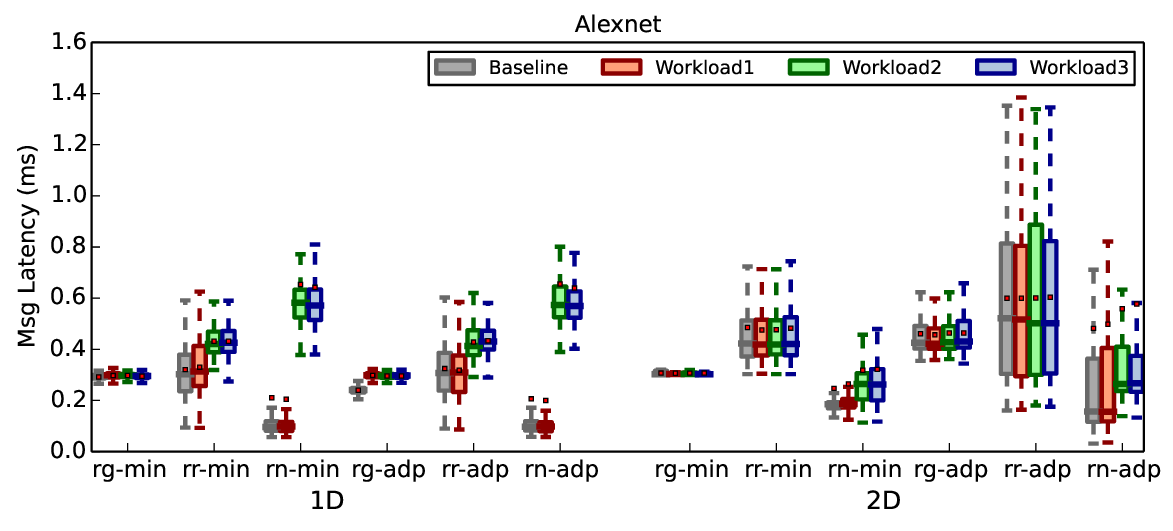}
}
\subfloat[Cosmoflow]{
   \includegraphics[width=0.48\textwidth]{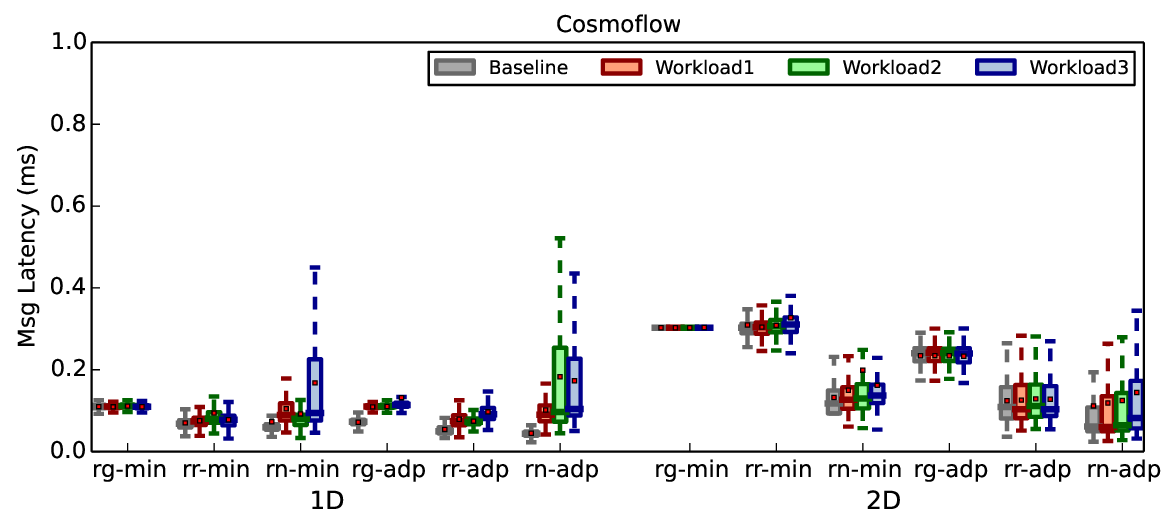}
}
\caption[]{Message Latency of each application with different configurations on 1D and 2D dragonfly systems. Different colors represent different workloads including baseline. Each box represents the minimum, first quartile, median, third quartile, and maximum, from bottom to top. The averages are shown in red squares.}
\label{fig:msgs}
\end{figure*}

\begin{figure*}[ht]
\centering
\subfloat[AlexNet with RR-ADP]{
   \includegraphics[width=0.42\textwidth]{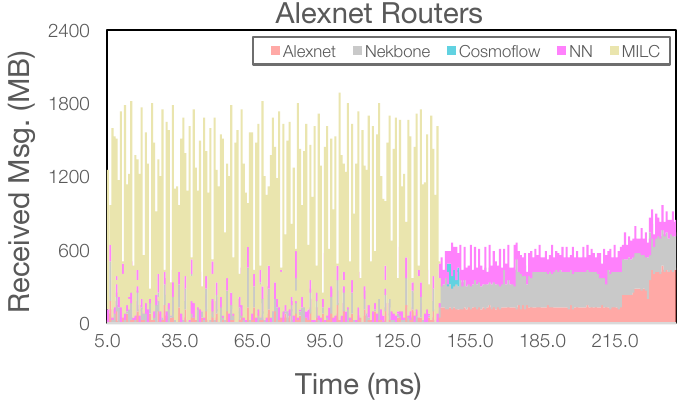}
}\hspace{0.08\textwidth}
\subfloat[AlexNet with RG-ADP]{
   \includegraphics[width=0.42\textwidth]{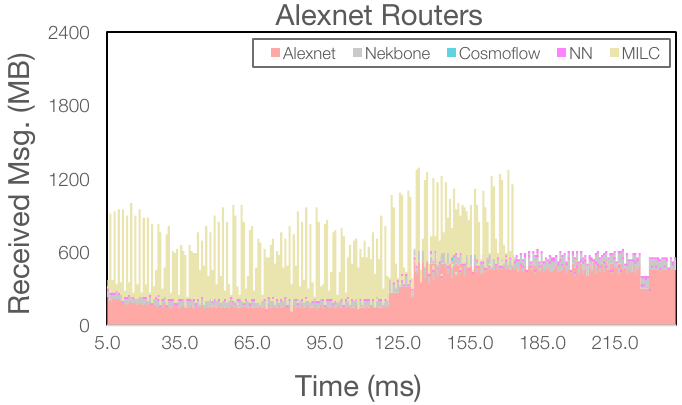}
}
\caption[]{The sum of messages received by all the routers that serve for AlexNet in Workload3 on 1D dragonfly network. Different colors represent different applications. }
\label{fig:counters}
\end{figure*}

In addition, we also assess the communication pattern by checking whether the data transmitted by each MPI rank match.
Table \ref{tab:correct2} shows the data transmitted in bytes by each rank for both the skeleton and the application.
The result demonstrates that the skeleton exhibits the same communication pattern to the corresponding application with each rank transmitting the same amount of bytes. 

Together, the above results indicate that the Union skeleton exhibits the same communication behaviors as the corresponding coNCePTuaL application.

\section{Hybrid Workload Analysis}
\label{sec:results}

\subsection{At Message Level}

Figure \ref{fig:msgs} compares the maximum message latency distributions for each application as shown in Table \ref{tab:workloads}, with different placement and routing combinations on two dragonfly systems. The distributions of message latency are shown as boxplots.
The baseline results shown in grey boxes indicate the ideal communication performance of the application when it has exclusive access to the system. 
We find that most applications have lower message latency with the random node placement than with other placement methods. 
Comparing different workloads with baseline performance under the same placement and routing configurations, we observe a maximum of 63x and 28x average latency slowdown for LAMMPS on 1D and 2D systems respectively.
For LAMMPS and Nekbone, the average message latency increases dramatically with the increase of workloads' communication intensity.
For MILC, the average latency delays are less than 11\% except 1.3x slowdown for 1D dragonfly system using adaptive routing.
For AlexNet and Cosmoflow, random router placement works better on 2D dragonfly than on 1D dragonfly. 
The average message latency delays with random router placement are within 2\% for all workloads on 2D dragonfly. 
Random node placement, on the other hand, causes more slowdown on both networks, with a maximum of 2x average latency delay on 1D dragonfly and a maximum of 24\% delay on 2D dragonfly, compared with the baseline cases.

Overall, we observe that the average message latency increases with the increase of the workloads' communication intensity. 
Communication intensive applications such as AlexNet and MILC, suffer less message latency delay than communication non-intensive applications such as LAMMPS and Nekbone. 
Across all workloads and placement/routing combinations, the maximum message latency delays are always observed with the random node placement, indicating that communication intensive jobs exacerbate the message latency of communication non-intensive jobs if they share the same network groups.
Confining messages within groups using random group placement helps reduce the network interference for both 1D and 2D dragonfly networks.


From the previous analysis, we observe that random group placement results in the smallest message latency delays for most of the cases. 
An explanation is that with random group placement, a job's messages are mostly confined within the assigned groups and could not easily interfere other jobs in different groups. 
To prove this point, we collect the time series data for messages on routers, and cluster the routers into different sets. 
Routers that connect to the nodes belong to the same job are grouped into one set. 
Therefore, by comparing the link traffic, we observe the amount of traffic each job's routers handled for other jobs.

Figure \ref{fig:counters} presents the total received messages on routers that serve AlexNet in Workload3 under random group and random router placement using adaptive routing. Messages from different applications are shown in different colors. 
The random node placement is omitted because one router may serve multiple jobs, which makes the per-router traffic data meaningless. 
The AlexNet routers handle an peak of 1800 MB messages from MILC, Nekbone and NN under the random router placement, whereas receiving only 800 MB messages from them under the random group placement. As a result, the AlexNet messages are received by its routers with a lower rate under the random router placement compared with that under the random group placement. 
This explains the phenomenon that AlexNet suffers from great message latency delays the under random router placement.
When jobs are allocated randomly at router level, sharing groups with communication intensive applications leads to link congestion and slows down their message arrival rate on the routers.
Separating jobs with random group placement helps reduce the messages from other job, and thus maintain a stable message arrival rate. 
Similar observations are found in other applications.

\begin{figure*}[ht]
\centering
\subfloat[LAMMPS]{
   \includegraphics[width=0.32\textwidth]{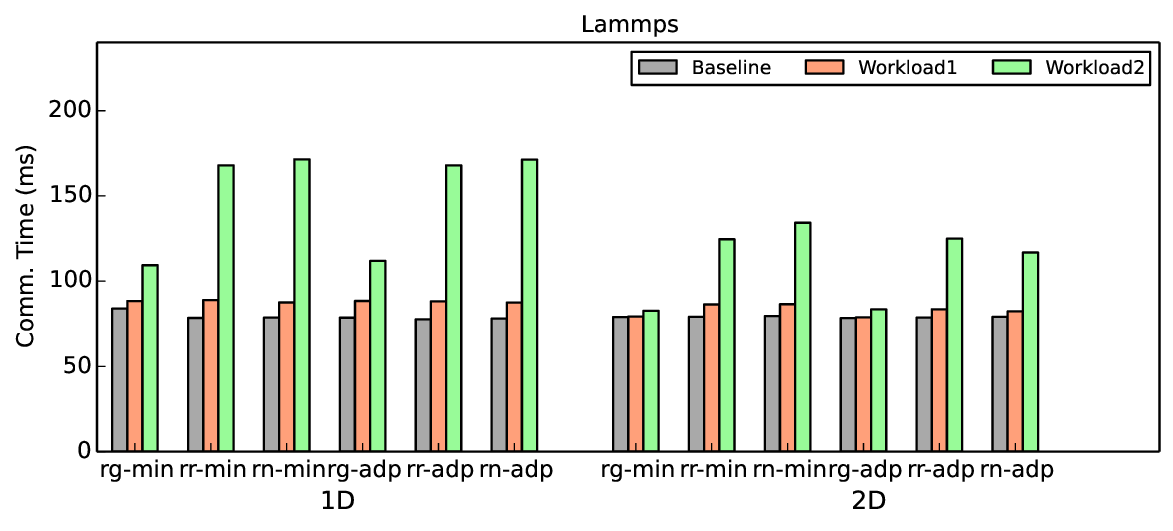}
}
\subfloat[Nekbone]{
   \includegraphics[width=0.32\textwidth]{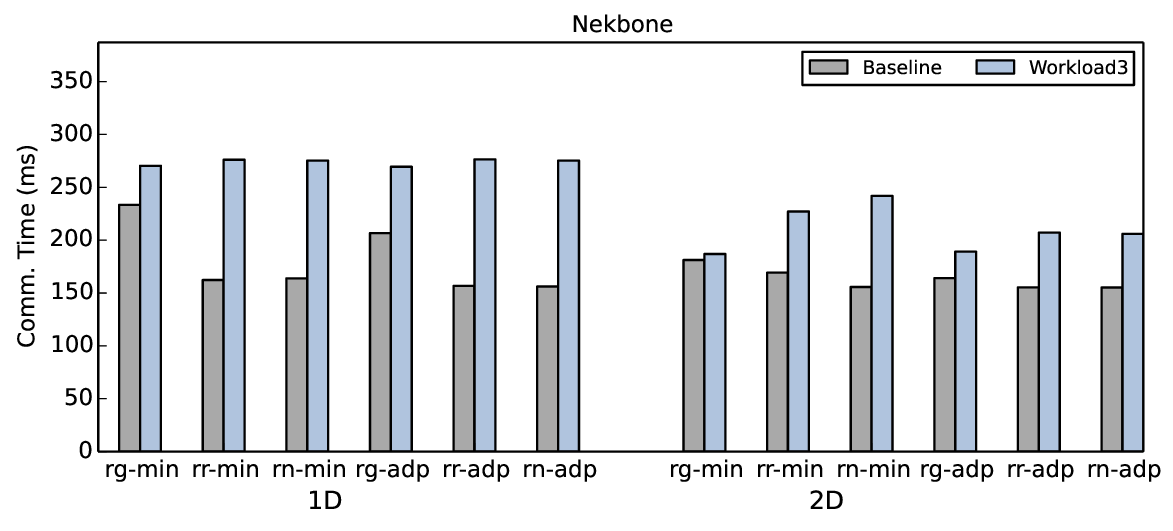}
}
\subfloat[MILC]{
   \includegraphics[width=0.32\textwidth]{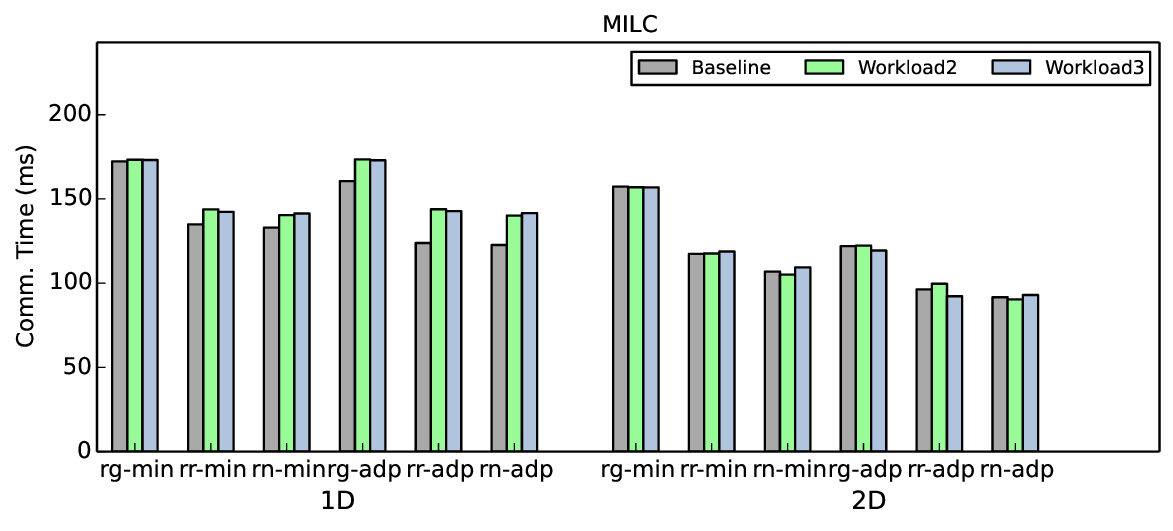}
}\\
\vspace{-8pt}
\subfloat[AlexNet]{
   \includegraphics[width=0.48\textwidth]{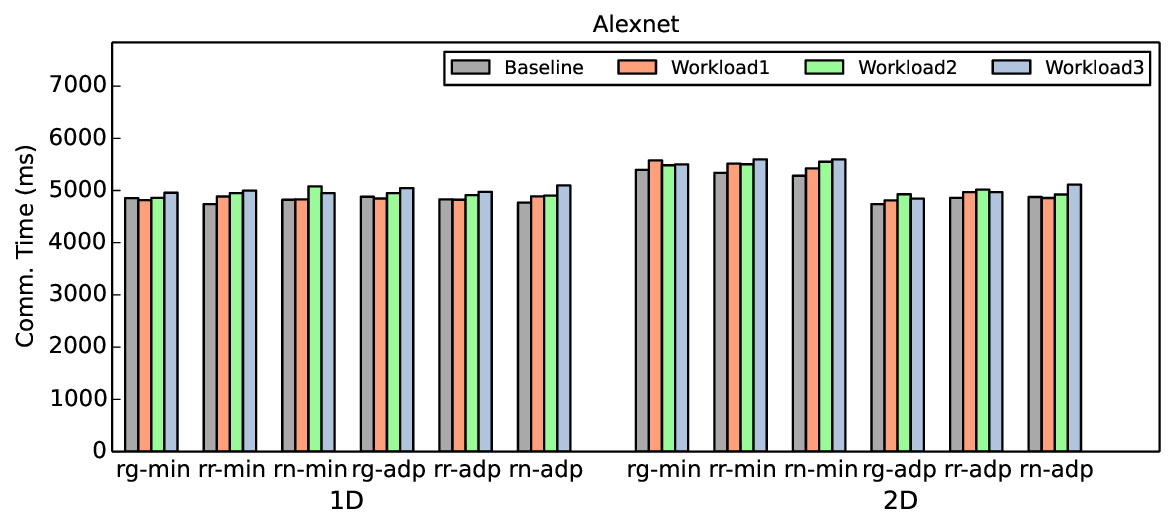}
}
\subfloat[Cosmoflow]{
   \includegraphics[width=0.48\textwidth]{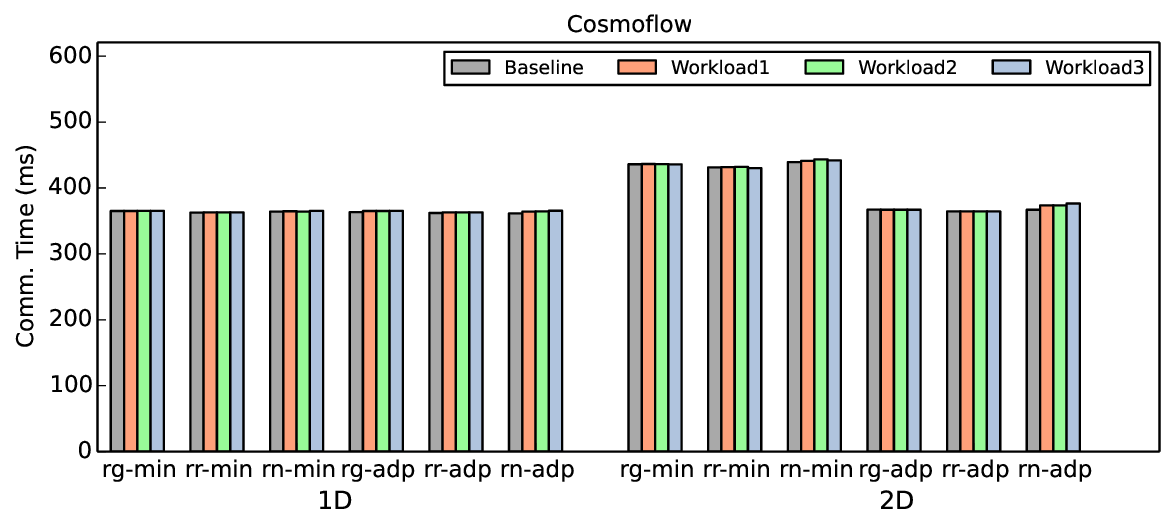}
}
\caption[]{Communication times of for each application with different configurations on 1D and 2D dargonfly systems. Different colors represent different workloads including baseline.}
\label{fig:comm}
\end{figure*}

\subsection{At Application Level}
Figure \ref{fig:comm} shows the comparisons of the maximum communication time for each application with different job placement and routing configurations on 1D and 2D dragonfly systems. Different colors represent different workloads.

For HPC applications, the baseline results show that all of them achieve better performance with random router/node placement. 
When the network is shared by multiple applications, random group placement helps reduce the network interference by confining most application traffic within the groups. 
Random group placement works better on 2D dragonfly than on 1D dragonfly.
LAMMPS and Nekbone that communicate with small or middle size messages are more sensitive to network interference, whereas MILC experiences less communication slowdown. 
Overall, the delays in message latency are reflected in the observed communication slowdowns for all three applications.
MILC is resistant to network interference, with the fact that it has a higher communication intensity compared with the other applications in the hybrid workloads. 

For ML applications, the baseline results show that changing job placement and routing does not have significant effect in communication time for both AlexNet and Cosmoflow on 1D dragonfly.
On 2D dragonfly, minimal routing leads to 15\% and 19\% communication time slowdown for AlexNet and Cosmoflow respectively, compared with adaptive routing.
Comparing between different networks, both AlexNet and Cosmoflow take longer time for communication on 2D dragonfly with minimal routing. 
For Cosmoflow, the communication time on 2D dragonfly can be 18.9\% greater than that on 1D dragonfly. 
Comparing between different workloads, however, we can not observe obvious communication slowdown for both applications. 
The significant delays in the message latency does not imply huge delays in the overall communication time. 
For example, the observed massage latency delays of AlexNet is 200\% in the random node placement with adaptive routing on 1D dragonfly, but the corresponding communication time slowndown is only 6.88\%. 
And there is no obvious difference in communication time for Cosmoflow between different workloads on both networks, though the maximum message latency delay reaches up to 24\%. 

Overall, the communication performance is not always consistent with the message latency performance. The increase in the message latency affects HPC applications more than ML applications. 
This indicates that ML applications can absorb the message latency variation better since they are featured with super-intensive blocking Allreduces.


\begin{table}[h]
\centering
\captionsetup{justification=centering}
\caption{Global and Local Link Load (Glink stands for global link, Llink stands for local link)}
\label{tab:linkload}
\begin{tabular}{c|l|l|l|l}
\hline\hline
\multicolumn{1}{c|}{Dragonfly} & \multicolumn{1}{l|}{\begin{tabular}[l]{@{}l@{}}Glink \\Load (TB)\end{tabular}} & \multicolumn{1}{l|}{\begin{tabular}[l]{@{}l@{}}Llink \\Load (TB)\end{tabular}} & \multicolumn{1}{l|}{\begin{tabular}[l]{@{}l@{}}Glink Load\\per link (MB)\end{tabular}} & \multicolumn{1}{l}{\begin{tabular}[l]{@{}c@{}}Llink Load\\per link (MB)\end{tabular}} \\ \hline
1D                 & 1.26                        & 5.33                 & 313.23                        & 5639.26                                                                              \\ 
2D                & 0.92                         & 10.01               & 65.39                          & 3214.65                                                                              \\ \hline\hline
\end{tabular}
\end{table}

\subsection{At System Level}

In theory, 3-hop 1D dragonfly should give a better network performance than 5-hop 2D dragonfly due to a smaller network diameter.
Because in a 2D dragonfly group, routers in different rows or columns don't have direct links between each other, packets need to traverse more hops to change dimensions compared with the all-to-all intra-group connected 1D dragonfly. 
However, in previous analysis, all HPC applications achieve smaller message latency and communication time on 2D dragonfly than on 1D dragonfly.

As shown in Table \ref{tab:sys}, the difference in group size and group number with different number of local and global links makes 2D dragonfly perform better than 1D dragonfly in this study. 
To illustrate this, we collect the end-of-simulation traffic information for each router port from Workload3 with random group placement and adaptive routing. We then calculate the sum of traffic on local links and global links of the whole system, and compute the average link load by dividing with the total number of links. The results are shown in Table \ref{tab:linkload}. 

1D dragonfly has smaller group size than 2D dragonfly, thus more traffic will be routed through global links. On 1D dragonfly, 19\% of the total traffic are routed through global links, compared with 8\% of that on 2D dragonfly. 
On the 1D dragonfly system, on average, each local and global link transmits more data than on the 2D dragonfly, hence  being more saturated and resulting in a higher message latency and a greater communication time as observed in previous analysis.

%

\subsection{Summary}
Application communication performance depends on many factors, including communication pattern, job placement, routing, and traffic on the shared network.
On a shared network, different job placement and routing mechanisms can lead to different traffic distributions. Bandwidth contention happens at the hot-spots, which leads to prolonged message latency. By carefully selecting job placement and routing mechanisms, we may achieve better application performance against network interference on a shared network like dragonfly.  

In summary, we have made several key findings. First, the results show that adaptive routing performs better than minimal routing under the same placement method, which is expected since adaptive routing is designed to avoid hot-spots and balance network traffic.

Second, the message latency is a reliable metric to reflect network interference. 
Application with intensive communication patterns suffers less message latency slowdown than communication non-intensive ones. 
Placing communication-intensive applications into separate groups helps confine their messages within the assigned groups, hence mitigating its interference to other applications. 

Third, the application communication performance is not always consistent with the message latency variation. The increase in the message latency affects HPC applications more than ML applications in term of communication time, implying that the ML application has a better ability to absorb the message delays.

Finally, in our system setup, applications achieve better performance on 2D dragonfly than on 1D dragonfly because 2D dragonfly offers more global and local links. With fewer links, 1D dragonfly has to handle higher traffic load per link, which makes the applications congest network more easily, resulting in delay of messages and slowdown of communication.


\section{Discussion}
In this study, we have focused on investigating the communication performance of co-running HPC and ML workloads on large-scale systems.
However, the convergence of ML workloads brings new challenges to the system design such as storage. 
ML applications usually require read-intensive I/O of a larger number of small files that need to be accessed in real-time during the training phases, putting large pressure on the storage system. 
We expect two major changes are needed in terms of introducing storage and I/O. 
First is at the application level where coNCePTuaL and Union will be enhanced to support I/O operations. 
Second is at the simulation level. We will leverage and extend the existing CODES storage module for concurrently simulating both communication and I/O traffics.
Building storage and I/O models for hybrid workload analysis is part of our future research.

Modeling and simulating hybrid workloads rely on application tracing that captures computation, communication, and I/O information. 
Many performance tools are available for such a purpose. 
For instance, we can use the open source DUMPI toolkit to collect MPI communication traces \cite{dumpi}, Darshan to capture the I/O access pattern for each process and file access pattern \cite{snyder2016modular}, and a PIN-based tracing tool to trace memory operations \cite{luk2005pin}.

This study aims to provide insights about how to optimize communication performance for different types of applications, by choosing appropriate job placement and routing mechanisms. 
For instance, the job placement findings presented in Section \ref{sec:results} could be used by batch schedulers 
for selecting appropriate job allocation, whereas the routing results could be used by a runtime system for changing message routes to mitigate network interference.


\section{Related Work}
\label{Sec:RelatedWork}


One promising approach to build cost-efficient large-scale simulation frameworks is to skeletonize or abstract applications so that only the execution flows remain but kernel computations are omitted for reducing the execution time. 
Jain et al. \cite{jain2016evaluating} use proxy applications as a simplified version of the original applications to evaluate HPC networks via parallel workload simulation. However, the development of new proxy applications is not easy.
Semi-automatic approaches 
\cite{wilke2018compiler} are proposed for extracting program skeletons based on compiler program analysis with the aid of user-provided annotations. This approach can be extended to large-scale simulations, but requires users to know where and how to annotate the original applications.
Our work, Union, helps make the large-scale parallel workload studies more convenient: i) simple and easy development of new applications by leveraging coNCepPTuaL, ii) automatic generation of skeletons, and iii) seamlessly integration with simulator.

Many studies have been conducted to explore the network interference in a multi-application environment on dragonfly topology. 
Chunduri et al.\cite{chunduri2017run} unveil the run-to-run job performance variation due to network interference on a production system.
Studies in \cite{Yang:SC:2016}, \cite{wang2018trade} explore the inter-application interference and suggest mitigating approaches. 
De Sensi et al.\cite{de2019mitigating} extract the performance counter information for network interference estimation and proposed an application-aware routing approach to improve performance.

To our knowledge, this is the first study for understanding performance implications of co-running scientific applications with machine learning applications on dragonfly systems.
We believe the key findings from hybrid workload analysis provide valuable insights for the HPC community.

\section{Conclusion}
\label{Sec:Conclusion}

We are heading towards the exascale computing era coupled with big data analytics using machine learning, understanding the performance implications of co-running scientific applications with big data and learning applications on extreme-scale systems are crucial for both system and application design. 

In this paper, we have presented Union as a workload manager that provides an automatic framework for generating in situ workloads and integrating these workloads in the network modeling toolkit CODES.
Union provides a unified and scalable workload management for large-scale network simulation. 
It significantly accelerates simulation by co-runing light-weight skeleton applications instead of traces. 
Union is available to the community as an open-source tool \cite{union}.

By using Union, we are able to conduct large-scale simulation studies of various hybrid workloads composed of traditional HPC applications and emerging ML applications. 
Our key findings show that the communication performance is not always consistent with the message latency performance. Network interference on HPC applications is more reflected by the message latency variation, whereas ML application has a better ability to absorb the message latency variation.


\section*{Acknowledgments}
This work is supported in part by US National Science Foundation grants CNS-1717763 and CCF-1618776.

\bibliography{mybib.bib}{}

\begin{thebibliography}{10}
\providecommand{\url}[1]{#1}
\csname url@samestyle\endcsname
\providecommand{\newblock}{\relax}
\providecommand{\bibinfo}[2]{#2}
\providecommand{\BIBentrySTDinterwordspacing}{\spaceskip=0pt\relax}
\providecommand{\BIBentryALTinterwordstretchfactor}{4}
\providecommand{\BIBentryALTinterwordspacing}{\spaceskip=\fontdimen2\font plus
\BIBentryALTinterwordstretchfactor\fontdimen3\font minus
  \fontdimen4\font\relax}
\providecommand{\BIBforeignlanguage}[2]{{%
\expandafter\ifx\csname l@#1\endcsname\relax
\typeout{** WARNING: IEEEtran.bst: No hyphenation pattern has been}%
\typeout{** loaded for the language `#1'. Using the pattern for}%
\typeout{** the default language instead.}%
\else
\language=\csname l@#1\endcsname
\fi
#2}}
\providecommand{\BIBdecl}{\relax}
\BIBdecl

\bibitem{ben2018demystifying}
T.~Ben-Nun and T.~Hoefler, ``Demystifying parallel and distributed deep
  learning: An in-depth concurrency analysis,'' \emph{arXiv preprint
  arXiv:1802.09941}, 2018.

\bibitem{you2018imagenet}
Y.~You, Z.~Zhang, C.-J. Hsieh, J.~Demmel, and K.~Keutzer, ``Imagenet training
  in minutes,'' in \emph{Proceedings of the 47th International Conference on
  Parallel Processing}.\hskip 1em plus 0.5em minus 0.4em\relax ACM, 2018, p.~1.

\bibitem{mathuriya2018cosmoflow}
A.~Mathuriya, D.~Bard, P.~Mendygral, L.~Meadows, J.~Arnemann, L.~Shao, S.~He,
  T.~K{\"a}rn{\"a}, D.~Moise, S.~J. Pennycook \emph{et~al.}, ``Cosmoflow: using
  deep learning to learn the universe at scale,'' in \emph{SC18: International
  Conference for High Performance Computing, Networking, Storage and
  Analysis}.\hskip 1em plus 0.5em minus 0.4em\relax IEEE, 2018, pp. 819--829.

\bibitem{kim2008technology}
J.~Kim, W.~J. Dally, S.~Scott, and D.~Abts, ``Technology-driven,
  highly-scalable dragonfly topology,'' in \emph{ACM SIGARCH Computer
  Architecture News}, vol.~36, no.~3.\hskip 1em plus 0.5em minus 0.4em\relax
  IEEE Computer Society, 2008, pp. 77--88.

\bibitem{faanes2012cray}
G.~Faanes, A.~Bataineh, D.~Roweth, E.~Froese, B.~Alverson, T.~Johnson,
  J.~Kopnick, M.~Higgins, J.~Reinhard \emph{et~al.}, ``Cray {Cascade}: {A}
  scalable {HPC} system based on a dragonfly network,'' in \emph{Proceedings of
  the International Conference on High Performance Computing, Networking,
  Storage and Analysis}.\hskip 1em plus 0.5em minus 0.4em\relax IEEE Computer
  Society Press, 2012, p. 103.

\bibitem{flajslik2018megafly}
M.~Flajslik, E.~Borch, and M.~A. Parker, ``Megafly: A topology for exascale
  systems,'' in \emph{International Conference on High Performance
  Computing}.\hskip 1em plus 0.5em minus 0.4em\relax Springer, 2018, pp.
  289--310.

\bibitem{chunduri2017run}
S.~Chunduri, K.~Harms, S.~Parker, V.~Morozov, S.~Oshin, N.~Cherukuri, and
  K.~Kumaran, ``Run-to-run variability on {Xeon} {Phi} based {Cray} {XC}
  systems,'' in \emph{SC17: International Conference for High Performance
  Computing, Networking, Storage and Analysis}.\hskip 1em plus 0.5em minus
  0.4em\relax IEEE, 2017.

\bibitem{jokanovic2015quiet}
A.~Jokanovic, J.~C. Sancho, G.~Rodriguez, A.~Lucero, C.~Minkenberg, and
  J.~Labarta, ``Quiet neighborhoods: {Key} to protect job performance
  predictability,'' in \emph{IEEE International Parallel and Distributed
  Processing Symposium (IPDPS)}.\hskip 1em plus 0.5em minus 0.4em\relax IEEE,
  2015, pp. 449--459.

\bibitem{kambadur2012measuring}
M.~Kambadur, T.~Moseley, R.~Hank, and M.~A. Kim, ``Measuring interference
  between live datacenter applications,'' in \emph{Proceedings of the
  International Conference on High Performance Computing, Networking, Storage
  and Analysis}.\hskip 1em plus 0.5em minus 0.4em\relax IEEE Computer Society
  Press, 2012, p.~51.

\bibitem{carothers2002ross}
C.~D. Carothers, D.~Bauer, and S.~Pearce, ``{ROSS}: {A} high-performance,
  low-memory, modular time warp system,'' \emph{Journal of Parallel and
  Distributed Computing}, vol.~62, no.~11, pp. 1648--1669, 2002.

\bibitem{mubarak2017enabling}
M.~Mubarak, C.~D. Carothers, R.~B. Ross, and P.~Carns, ``Enabling parallel
  simulation of large-scale {HPC} network systems,'' \emph{IEEE Transactions on
  Parallel and Distributed Systems}, vol.~28, no.~1, pp. 87--100, 2017.

\bibitem{rodrigues2011structural}
A.~F. Rodrigues, K.~S. Hemmert, B.~W. Barrett, C.~Kersey, R.~Oldfield,
  M.~Weston, R.~Risen, J.~Cook, P.~Rosenfeld, E.~CooperBalls \emph{et~al.},
  ``The structural simulation toolkit,'' \emph{SIGMETRICS Performance
  Evaluation Review}, vol.~38, no.~4, pp. 37--42, 2011.

\bibitem{jiang2013detailed}
N.~Jiang, D.~U. Becker, G.~Michelogiannakis, J.~Balfour, B.~Towles, D.~E. Shaw,
  J.~Kim, and W.~J. Dally, ``A detailed and flexible cycle-accurate
  network-on-chip simulator,'' in \emph{2013 IEEE International Symposium on
  Performance Analysis of Systems and Software (ISPASS)}.\hskip 1em plus 0.5em
  minus 0.4em\relax IEEE, 2013, pp. 86--96.

\bibitem{thompson2014scalable}
\BIBentryALTinterwordspacing
J.~Thompson, ``Scalable workload models for system simulations,'' Intel, Tech.
  Rep., 08 2014. [Online]. Available:
  \url{http://hpc.pnl.gov/modsim/2014/Presentations/Thompson.pdf}
\BIBentrySTDinterwordspacing

\bibitem{jain2016evaluating}
N.~Jain, A.~Bhatele, S.~White, T.~Gamblin, and L.~V. Kale, ``Evaluating hpc
  networks via simulation of parallel workloads,'' in \emph{SC'16: Proceedings
  of the International Conference for High Performance Computing, Networking,
  Storage and Analysis}.\hskip 1em plus 0.5em minus 0.4em\relax IEEE, 2016, pp.
  154--165.

\bibitem{mubarak2019evaluating}
M.~Mubarak, N.~McGlohon, M.~Musleh, E.~Borch, R.~B. Ross, R.~Huggahalli,
  S.~Chunduri, S.~Parker, C.~D. Carothers, and K.~Kumaran, ``Evaluating quality
  of service traffic classes on the megafly network,'' in \emph{International
  Conference on High Performance Computing}.\hskip 1em plus 0.5em minus
  0.4em\relax Springer, 2019, pp. 3--20.

\bibitem{de2019mitigating}
D.~De~Sensi, S.~Di~Girolamo, and T.~Hoefler, ``Mitigating network noise on
  dragonfly networks through application-aware routing,'' \emph{arXiv preprint
  arXiv:1909.07865}, 2019.

\bibitem{pakin2007design}
S.~{Pakin}, ``The design and implementation of a domain-specific language for
  network performance testing,'' \emph{IEEE Transactions on Parallel and
  Distributed Systems}, vol.~18, no.~10, pp. 1436--1449, Oct 2007.

\bibitem{pakin2004conceptual}
S.~{Pakin}, ``{coNCePTuaL}: a network correctness and performance testing
  language,'' in \emph{18th International Parallel and Distributed Processing
  Symposium, 2004. Proceedings.}, April 2004, pp. 79--.

\bibitem{wang2018trade}
X.~{Wang}, M.~{Mubarak}, X.~{Yang}, R.~B. {Ross}, and Z.~{Lan}, ``Trade-off
  study of localizing communication and balancing network traffic on a
  dragonfly system,'' in \emph{2018 IEEE International Parallel and Distributed
  Processing Symposium (IPDPS)}, May 2018, pp. 1113--1122.

\bibitem{mubarak2017quantifying}
M.~Mubarak, P.~Carns, J.~Jenkins, J.~K. Li, N.~Jain, S.~Snyder, R.~Ross, C.~D.
  Carothers, A.~Bhatele, and K.-L. Ma, ``Quantifying i/o and communication
  traffic interference on dragonfly networks equipped with burst buffers,'' in
  \emph{IEEE International Conference on Cluster Computing (CLUSTER)}.\hskip
  1em plus 0.5em minus 0.4em\relax IEEE, 2017, pp. 204--215.

\bibitem{Yang:ICPADS:2016}
X.~Yang, J.~Jenkins, M.~Mubarak, X.~Wang, R.~B. Ross, and Z.~Lan, ``Study of
  intra- and interjob interference on torus networks,'' in \emph{IEEE 22nd
  International Conference on Parallel and Distributed Systems (ICPADS)}, Dec
  2016, pp. 239--246.

\bibitem{seo2017argobots}
S.~Seo, A.~Amer, P.~Balaji, C.~Bordage, G.~Bosilca, A.~Brooks, P.~Carns,
  A.~Castell{\'o}, D.~Genet, T.~Herault \emph{et~al.}, ``Argobots: A
  lightweight low-level threading and tasking framework,'' \emph{IEEE
  Transactions on Parallel and Distributed Systems}, vol.~29, no.~3, pp.
  512--526, 2017.

\bibitem{sergeev2018horovod}
A.~Sergeev and M.~D. Balso, ``Horovod: fast and easy distributed deep learning
  in {TensorFlow},'' \emph{arXiv preprint arXiv:1802.05799}, 2018.

\bibitem{Bebop}
\BIBentryALTinterwordspacing
\emph{Bebop at LCRC}. [Online]. Available:
  \url{http://www.lcrc.anl.gov/systems/resources/bebop}
\BIBentrySTDinterwordspacing

\bibitem{dumpi}
\BIBentryALTinterwordspacing
\emph{DUMPI}. [Online]. Available:
  \url{https://github.com/sstsimulator/sst-dumpi}
\BIBentrySTDinterwordspacing

\bibitem{snyder2016modular}
S.~Snyder, P.~Carns, K.~Harms, R.~Ross, G.~K. Lockwood, and N.~J. Wright,
  ``Modular hpc i/o characterization with darshan,'' in \emph{2016 5th Workshop
  on Extreme-Scale Programming Tools (ESPT)}.\hskip 1em plus 0.5em minus
  0.4em\relax IEEE, 2016, pp. 9--17.

\bibitem{luk2005pin}
C.-K. Luk, R.~Cohn, R.~Muth, H.~Patil, A.~Klauser, G.~Lowney, S.~Wallace, V.~J.
  Reddi, and K.~Hazelwood, ``Pin: building customized program analysis tools
  with dynamic instrumentation,'' in \emph{Acm sigplan notices}, vol.~40,
  no.~6.\hskip 1em plus 0.5em minus 0.4em\relax ACM, 2005, pp. 190--200.

\bibitem{wilke2018compiler}
J.~J. Wilke, J.~P. Kenny, S.~Knight, and S.~Rumley, ``Compiler-assisted
  source-to-source skeletonization of application models for system
  simulation,'' in \emph{International Conference on High Performance
  Computing}.\hskip 1em plus 0.5em minus 0.4em\relax Springer, 2018, pp.
  123--143.

\bibitem{Yang:SC:2016}
\BIBentryALTinterwordspacing
X.~Yang, J.~Jenkins, M.~Mubarak, R.~B. Ross, and Z.~Lan, ``Watch out for the
  bully!: {Job} interference study on dragonfly network,'' in \emph{Proceedings
  of the International Conference for High Performance Computing, Networking,
  Storage and Analysis}, ser. SC '16.\hskip 1em plus 0.5em minus 0.4em\relax
  Piscataway, NJ, USA: IEEE Press, 2016, pp. 64:1--64:11. [Online]. Available:
  \url{http://dl.acm.org/citation.cfm?id=3014904.3014990}
\BIBentrySTDinterwordspacing

\bibitem{union}
\BIBentryALTinterwordspacing
\emph{Union}. [Online]. Available: \url{https://github.com/SPEAR-IIT/Union.git}
\BIBentrySTDinterwordspacing

\end{thebibliography}
\bibliographystyle{IEEEtran}

\end{document}